\documentclass[conference]{IEEEtran}
\usepackage[top=0.75in,bottom=1.05in,left=0.625in,right=0.63in]{geometry}
\usepackage{caption}
\usepackage{lipsum}
\usepackage{geometry}
\usepackage{graphicx}
\usepackage{multicol}
\usepackage{diagbox}
\usepackage{blkarray}

\usepackage{hyperref}
\usepackage{float}
\usepackage{url}
\usepackage[utf8]{inputenc} 
\usepackage{array}
\usepackage[utf8]{inputenc} 
\usepackage[T1]{fontenc}
\usepackage{stfloats}
\usepackage{mathtools}
\usepackage{amsfonts}
\usepackage{tabstackengine}
\usepackage{amssymb}
\usepackage{enumerate}
\usepackage{cite}
\usepackage{calc}
\usepackage{nicematrix}
\usepackage{color}
\usepackage{epsfig}
\usepackage{setspace}
\usepackage{pstricks}
\usepackage{cancel}
\usepackage{multirow}
\usepackage{mathrsfs}
\usepackage{algorithm}
\usepackage{algpseudocode}
\usepackage{multirow}
\usepackage{romannum}
\usepackage{diagbox}
\algnewcommand{\algorithmicor}{\textbf{ or }}
\algnewcommand{\OR}{\algorithmicor}
\usepackage{booktabs,makecell}
\usepackage{mathrsfs}
\usepackage{enumitem}
\newtheorem{defn}{Definition}
\newtheorem{thm}{Theorem}
\newtheorem{cor}{Corollary}

\newtheorem{remark}{Remark}

\newtheorem{claim}{Claim}
\newtheorem{example}{Example}
\graphicspath{{./images/}}
\def\Ddots{\mathinner{\mkern1mu\raise\p@
		\vbox{\kern7\p@\hbox{.}}\mkern2mu
		\raise4\p@\hbox{.}\mkern2mu\raise7\p@\hbox{.}\mkern1mu}}
\makeatother

\title{Combinatorial Multi-Access Coded Caching with Private Caches under Intersecting Index Constraints}
					\author{\IEEEauthorblockN{Dhruv Pratap Singh}
						\IEEEauthorblockA{
							\textit{Dept. of ECE, Indian Institute of Science}\\
							Bangalore, KA, India \\
							dhruvpratap@iisc.ac.in}
						\and
						\IEEEauthorblockN{Anjana A. Mahesh}
						\IEEEauthorblockA{
							\textit{Dept. of EE, Indian Institute of Tech. }\\ 
						Hyderabad, Telangana, India \\
						anjana.am@ee.iith.ac.in}
					\and
					\IEEEauthorblockN{B. Sundar Rajan}
					\IEEEauthorblockA{
						\textit{Dept. of ECE, Indian Institute of Science}\\
						Bangalore, KA, India \\
						bsrajan@iisc.ac.in}}
\begin{document}
	\pagenumbering{arabic}
	\maketitle
	\begin{abstract}
		We consider the coded caching system introduced in “Combinatorial Multi-Access Coded Caching with Private Caches” by Singh et al., where each user, equipped with a private cache, accesses a distinct $r-$subset of access caches. A central server housing a library of files populates both private and access caches using uncoded placement. In this work, we focus on a constrained indexing regime, referred to as the intersection class, in which the sets used to index the demands of each user must have a nonempty intersection. This regime models resource-limited IoT scenarios such as edge-assisted IoT systems, where devices with small private caches connect to a small number of shared caches. 
		We provide a necessary and sufficient condition under which the system parameters fall within this intersection class. Under this condition, we propose a centralized coded caching scheme and characterize its rate-memory trade-off. Next, we define a uniform-intersection subclass and establish a condition under which the system belongs to this subclass. Within this subclass, the proposed scheme has a regular structure, with each transmission benefiting the same number of users, and we characterize its rate-memory trade-off. Additionally, we derive an index coding-based lower bound on the minimum achievable worst-case rate under uncoded placement. Finally, we provide numerical comparisons between the rate of the proposed scheme, the new lower bound, and bounds from the original work.
	\end{abstract}
	\section{Introduction}
	
	The rapid growth of demand for high-definition video streaming services has led to a significant surge in internet data traffic. To alleviate the resulting load on the communication infrastructure, Maddah-Ali and Niesen proposed the coded caching technique in their seminal work \cite{MAN}. Coded caching operates in two phases: the \textit{placement phase}, during which the server prefetches content into the system memories during off-peak hours, and the \textit{delivery phase}, where the server transmits coded multicast messages to fulfill user demands during peak hours. The objective in coded caching is to jointly design the placement and delivery strategies to minimize the number of transmissions during the delivery phase.

		The scheme proposed by Maddah-Ali and Niesen, referred to as the MAN scheme \cite{MAN}, considers a centralized server with $N$ files connected to $K$ users via a shared bottleneck wireless link. Each user possesses a cache of size $M \leq N$. The MAN scheme was proven optimal under uncoded placement for $N \geq K$ in \cite{WTP} and for $N < K$ in \cite{YMA}.	Subsequent research has extended the coded caching framework to various settings \cite{AN}-\cite{SMR}, among which users in \cite{SMR} access two different types of caches, one shared between multiple users and the other private to a user.

This work considers the Combinatorial Multi-Access with Private caches (CMAP) coded caching model introduced in \cite{SMR}, where a central server housing $N$ files connects to $K$ users and $\Lambda$ access caches via an error-free wireless broadcast link. Each user is equipped with a private cache, i.e., a local memory component that is exclusively accessible to that user, and connects to a distinct $r$-subset of the $\Lambda$ access caches.

Edge-assisted IoT networks increasingly rely on caching to reduce latency and alleviate core network load, but are typically constrained by limited memory at both edge nodes and end devices. Access caches, deployed at gateway nodes or edge servers, often have restricted storage due to cost, energy, and hardware limitations, while IoT devices such as sensors and embedded controllers possess minimal local memory for reasons of power efficiency and simplicity. The total number of access caches is also typically small in practice, as observed in smart buildings, where a few edge nodes serve a large number of devices. In such systems, each user connects to a subset of access caches, with the specific connectivity pattern determined by proximity, deployment topology, and network design. These characteristics align closely with the CMAP framework, which captures structured user-to-cache connectivity in environments featuring heterogeneous devices with constrained communication, computation, and caching capabilities \cite{TGH}. Unlike traditional host-centric architectures, IoT networks increasingly adopt information-centric paradigms that emphasize in-network caching to reduce latency and improve data availability \cite{NUMAL}. Efficient caching both at individual devices (private caches) and at shared network nodes (access caches), is critical for minimizing redundant transmissions, reducing communication overhead, and optimizing energy usage \cite{WCW}. 
These considerations motivate the study of a constrained regime of the CMAP system in this work, which captures practical limitations on cache resources and infrastructure commonly encountered in real-world IoT deployments.

While \cite{SMR} proposed a coded caching scheme for a fixed value of private cache memory, this work focuses on a constrained indexing regime, referred to as the intersection class, where the sets used to index the user-demands must have a nonempty intersection. Within this class, we propose a centralized scheme for which the private cache memory can take different values, as long as it satisfies the intersection class condition. We also identify a uniform-intersection subclass, where the cardinality of the intersection of the sets used to index the user-demands is uniform across all user-demands.

\textit{Organization of the paper:} Section \ref{systemmodelsection} introduces the system model along with the necessary preliminaries for the results presented later in the paper. In Section~\ref{mainresults}, we define the intersection class, identify a regime of parameter values for which the CMAP setting falls in this class, and present the rate of a proposed centralized coded caching scheme. We also introduce the uniform-intersection subclass, provide a condition on the system parameters under which the system falls within this subclass, and characterize the rate in this subclass. Next, we present a lower bound on the optimal worst-case rate under uncoded placement using index coding arguments. The proposed delivery scheme is described in Section~\ref{achievabilitysection}, and Section~\ref{numericalcomparisons} presents numerical comparisons with existing bounds from \cite{SMR} and the lower bound proposed in Section \ref{mainresults}. 

	\textit{Notations:} The set $\{1,2,\cdots,N\}$ is denoted as $[N]$, whereas the set $\{a,a+1,\cdots,b\}$ where $b\geq a, a \neq 1$ is denoted by $[a,b]$, for some $a,b\in\mathbb{Z}^+$, where $\mathbb{Z}^+$ is the set of all non-negative integers. The cardinality of a set $A$ is denoted as $|A|$. We use $A_{i,i\in[n]}$ to denote $\{A_1,A_2,\cdots,A_n\}$. 
	The binomial coefficient $\frac{n!}{k!(n-k)!}$ is denoted as $\binom{n}{k}$ and we assume $\binom{n}{k}=0$ if $n < 0, k < 0$ or $n<k$. We use the $\oplus$ symbol to denote the bitwise XOR operation.
	\section{System Model and Preliminaries}
	\label{systemmodelsection}
	In this section, we introduce the system model. We then revisit relevant bounds from \cite{SMR} and results from index coding that will be used in this work.
	\subsection{System Model}
	\begin{figure}
		\centering
		\includegraphics[height=0.37\columnwidth,width=0.97\columnwidth]{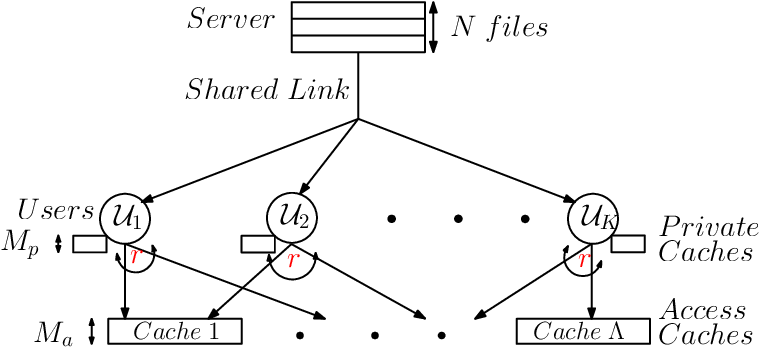}
		\caption{The $(\Lambda,r,M_a,M_p,N)-$CMAP Coded Caching System.}
		\label{systemmodel}
	\end{figure}
	Consider the system model shown in Fig. \ref{systemmodel}. A central server stores a library of $N$ files, $W_1, W_2, \cdots, W_N$, each of size $B$ bits. There are $\Lambda$ caches in the system, each with a storage capacity of $M_a$ files, where $M_a \leq N$. These $\Lambda$ caches are referred to as the access caches. In addition, there are users, each equipped with a private cache of size $M_p \leq N$, that connect to the access caches through an infinite capacity, error-free wireless link. Each user is connected to a distinct $r-$subset of the $\Lambda$ access caches, where $r$ is the access degree. Hence, each user is indexed by the set of access caches it is connected to, i.e., by a subset of $[\Lambda]$ of cardinality $r$. This topology, referred to as the combinatorial multi-access topology \cite{PD,BE}, results in $K = \binom{\Lambda}{r}$ users, where $K \leq N$. The server communicates with the 	users through a shared, error-free wireless bottleneck link. We are interested in memory points $(M_a, M_p)$ satisfying $0 < rM_a + M_p < N$. We refer to this system as the $(\Lambda, r, M_a, M_p, N)-$CMAP coded caching system. The CMAP coded caching system operates in two phases:
	\begin{enumerate}
		\item \textit{Placement Phase:} 
				The server populates both the private and access caches using an uncoded placement strategy. Specifically, each file $W_n$ in the library is first divided into subfiles as $W_n=\{W_{n,S}:S\subset[\Lambda],|S|=\frac{\Lambda M_a}{N}\}$, which are then stored across the access caches to populate them such that access cache $a$ stores the subfiles $W_{n,S}$, for all $n\in [N],$ if $a\in S$. Each subfile $W_{n,S}$ is further partitioned into mini-subfiles as $W_{n,S}=\{W_{n,S,T_{i,i\in\left[\frac{KM_p}{N}\right]}}:T_i\subset[\Lambda]\setminus S,|T_i|=r,\forall i\in\left[\frac{KM_p}{N}\right]\}$, where each mini-subfile $W_{n,S,T_{i,i\in\left[\frac{KM_p}{N}\right]}}$ is subsequently prefetched in the private caches of the users $T_{1},T_{2},\cdots,T_{\frac{KM_p}{N}}$. The total number of mini-subfiles each file is divided into is referred to as the subpacketization level and is denoted by $F$. The contents stored in access cache $a \in [\Lambda]$ are denoted by $Z_a,$ and the contents of the private cache of user $U \subseteq [\Lambda],|U|=r$ are denoted by $Z^p_U$. 
		\item \textit{Delivery Phase:} Each user $U \subseteq [\Lambda], |U| = r$ requests one file from the server. The collection of all user demands is represented by the demand vector $\mathbf{d} = (d_U : U \subseteq [\Lambda], |U| = r)$, where $d_U \in [N]$ denotes the index of the file requested by user $U$. We denote $W_{d_{U},S,T_{i,i\in\left[\frac{KM_p}{N}\right]}}$ as the mini-subfile corresponding to subfile $S$, which is stored in the private cache of users $T_1,T_2,\cdots,T_{\frac{KM_p}{N}}$ and is demanded by user $U$. Upon receiving the demand vector $\mathbf{d}$, the server transmits a sequence of coded messages, each comprising of coded combinations of the mini-subfiles, to satisfy the demands of the users. The objective is to minimize the number of bits transmitted by the server. The rate of a scheme is defined as the total number of bits transmitted by the server, normalized by the file size $B$. The worst-case rate corresponds to the scenario where all users demand distinct files.
	\end{enumerate} 
	\begin{defn}
Consider a $(\Lambda, r, M_a, M_p, N)$-CMAP coded caching system. A rate-memory triplet $(M_a, M_p, R)$ is said to be achievable if there exists a coded caching scheme, under any general placement, that achieves rate $R$ at memory point $(M_a, M_p)$, for sufficiently large file size $B$, under any demand vector. The optimal worst-case rate for the $(\Lambda, r, M_a, M_p, N)$-CMAP system is defined as
\begin{align}
R^{\textasteriskcentered}(M_a, M_p) = \inf \left\{ R : (M_a, M_p, R) \text{ is achievable} \right\},
\end{align}
	\end{defn}The server seeks to design jointly the placement and delivery policies to achieve $R^{\textasteriskcentered}_{M_a,M_p}$.
	\subsection{Some Bounds discussed in \cite{SMR}}
	\label{Preliminarybounds}
	In this subsection, we review 
	certain bounds from \cite{SMR} that will be employed in the Section \ref{numericalcomparisons} of this paper. 
	
		
	
	
	\begin{thm}
		\label{prop1}
		For a $(\Lambda,r, M_a, M_p, N)-$CMAP coded caching system, the optimal worst-case rate $R_{UC}^{\textasteriskcentered}(M_a, M_p)$ under uncoded placement is bounded as
			$R^{\textasteriskcentered}_{D}(rM_a+M_p)\leq R_{UC}^{\textasteriskcentered}(M_a,M_p)\leq R^{\textasteriskcentered}_{CMACC}(M_a+\frac{M_p}{r}),$
		where, $R^{\textasteriskcentered}_{D}({rM_a+M_p})$ is the rate achieved by MAN scheme \cite{MAN} {by letting $M=rM_a+M_p$} and $R^{\textasteriskcentered}_{CMACC}({M_a+\frac{M_p}{r}})$ is the rate achieved by MAN scheme for CMACC network \cite{PD} {by letting $M=M_a+\frac{M_p}{r}$}.
	\end{thm} 
	\begin{thm}
		\label{thm1}
		For a $(\Lambda,r,M_a,M_p,N)-$CMAP coded caching system, the worst-case rate is lower bounded as
			$R^{\textasteriskcentered}(M_a,M_p)\geq\max\limits_{s\in\{1,2,\cdots,K\}} \left(s-\frac{qM_a+sM_p}{\left\lfloor\frac{N}{s}\right\rfloor}\right),$
		where $q=\min\{\Lambda+r-1,\Lambda\}$.\end{thm}
	\subsection{Index Coding}
	The index coding problem (ICP) with side information \cite{BK}, \cite{YBJK} involves a single source that has $ n $ messages $ x_1, x_2, \dots, x_n $, where $ x_i \in \mathbb{F}_q $ for all $ i \in [n] $, and broadcasts to a set of $ K $ receivers $ R_1, R_2, \dots, R_K $. Each receiver $ R_{i}: i \in [K] $, has prior knowledge of a subset $ \{x_j : j \in {X}_i\} $, where $ {X}_i \subseteq [n] $ denotes its side information. Further, receiver $ R_i $ demands message $ x_{f(i)} $, where $ f : [K] \rightarrow [n] $ and $ f(i) \notin {X}_i $.
	
	For an instance $ \mathcal{I} $ of the ICP, the generalized independence number $ \alpha(\mathcal{I}) $ was defined in \cite{DSC} as follows: Let $ \mathcal{Y}_i = [n] \setminus (\{f(i)\} \cup {X}_i) $ for each receiver $ R_i $, and define the collection $ \mathcal{J}(\mathcal{I}) = \bigcup_{i \in [K]} \{\{f(i)\} \cup Y_i : Y_i \subseteq \mathcal{Y}_i\} $. A subset $ H \subseteq [n] $ is called a generalized independent set if every subset of $ H $ belongs to $ \mathcal{J}(\mathcal{I}) $. The largest such set is referred to as the maximal generalized independent set, and its cardinality, denoted by $ \alpha(\mathcal{I}) $, is called the generalized independence number. It was shown in \cite{KTR} that $ \alpha(\mathcal{I}) $ provides a lower bound on the number of scalar linear transmissions required to solve the ICP $ \mathcal{I} $.
	
	In the context of coded caching, for a given uncoded placement scheme and demand vector, each user can be modeled as a receiver in an ICP, where the desired message comprises of the subfiles of the requested file, and the side information consists of the subfiles available in the user’s cache. Therefore, the delivery phase of the coded caching problem can be formulated as an instance of the ICP, and the corresponding generalized independence number provides a lower bound on the number of scalar linear transmissions required to satisfy all user demands \cite{KTR}.
	
	%
	
	\section{Main Results}
	\label{mainresults}
	
We begin this section by introducing the intersection class in Definition~\ref{intersectionclassdefn}. Claim~\ref{intersectionclassclaim} provides a necessary and sufficient condition under which a $(\Lambda, r, M_a, M_p, N)-$CMAP system belongs to the intersection class. Theorem~\ref{intersectionclassrate} establishes an achievable rate for this class. We then define the uniform-intersection subclass in Definition~\ref{uniformintersectiondefn}, and in Claim~\ref{uniformintersectionclaim}, we identify a condition under which a system belongs to this subclass. Corollary~\ref{uniformintersectionrate} characterizes the rate-memory trade-off for the uniform-intersection subclass. Finally, Theorem~\ref{alphaboundrate} presents a lower bound on the optimal worst-case rate under the proposed uncoded placement strategy, as described in Section \ref{achievability}.
	
	
	\begin{defn}
		\label{intersectionclassdefn}
		A $(\Lambda,r,M_a,M_p,N)-$CMAP coded caching system belongs to the intersection class if the cardinality of the intersection of the user-index set $U$ and the mini-subfile-index sets $T_{i,i\in[t_p]}$ is non-empty for all mini-subfiles. Specifically for any mini-subfile $W_{d_{{U}},{S},{T}_{i,i\in[t_p]}},$ the intersection set $I=\{\bigcap\limits_{i=1}^{t_p} {T}_i\}\cap U\not=\emptyset$, 
		where $t_p=\frac{K M_p}{N}$.
	\end{defn}
	\begin{claim} 
		\label{intersectionclassclaim}
		For any mini-subfile $W_{d_{{U}}, {S}, {T}_{i\in[t_p]}}$, the intersection set $I = \{\bigcap\limits_{i=1}^{t_p} {T}_i\}\cap U \neq \emptyset$ if and only if $r>\frac{\Lambda(1-\frac{M_a}{N})}{1+\frac{N}{K M_p}}$.
	\end{claim}
	\begin{IEEEproof}
	\label{intersectionclassclaimproof}
	Observe that $r>\frac{\Lambda(1-\frac{M_a}{N})}{1+\frac{N}{K M_p}}$ can be written as $\Lambda< t_a+r+\frac{r}{t_p}$, where $t_a=\frac{\Lambda M_a}{N}$ and $t_p=\frac{KM_p}{N}$. We use this form of the condition to prove the claim.
	
	We prove this claim in two steps. First, we prove that for $\Lambda < t_a + r+\frac{r}{t_p}$, there exists no mini-subfile such that the intersection set $I=\emptyset$. This is proven via contradiction. Following this, we show that for any $\Lambda \geq t_a + r+\frac{r}{t_p}$, mini-subfiles with $I=\emptyset$ will always exist.
		We now proceed with the first part of the proof.
		\begin{enumerate}
			\item \textit{Proof by Contradiction}: Assume that for $\Lambda < t_a + r+\frac{r}{t_p}$, there exists a user-index set ${U}$ and mini-subfile-index sets ${T}_{i,i\in[t_p]}$ such that their intersection is empty, i.e., the intersection set $ I = \{\bigcap\limits_{i=1}^{t_p} {T}_i\}\cap U = \emptyset $. Each set ${U}$ and ${T}_{i,i \in [t_p]}$ has cardinality $r$ and is a subset of $ [\Lambda] \setminus {S} $, where ${S} \subset [\Lambda]$ and $|{S}| = t_a$. 
			
			For each integer $ j \in [\Lambda] \setminus {S} $ appearing in these sets, let $ i_j $ denote its number of occurrences across the $ (t_p+1) $ sets, ${U}$ and ${T}_{i,i\in[t_p]}$. 
			%
			It follows that $i_j\leq t_p$ because if $i_j=t_p+1$, then $I\not=\emptyset$, contradicting our assumption. Given that $j\in[\Lambda]\setminus{S}$, we have \begin{align*}\sum\limits_{j\in[\Lambda]\setminus{S}} i_j\leq (\Lambda-t_a)t_p.\end{align*} But from the claim, we have $(\Lambda-t_a)t_p< (t_p+1)r$, leading to \begin{align*}\sum\limits_{j\in[\Lambda]\setminus{S}} i_j<(t_p+1)r.\end{align*}This is a contradiction as each set has cardinality $r$, so the sum of occurrences of all the integers $\sum\limits_{j\in[\Lambda]\setminus{S}} i_j = (t_p+1)r$. Hence, for $\Lambda < t_a + r + \frac{r}{t_p}$, there exists no mini-subfile such that the intersection set $I=\emptyset$. 
			
			We now present the second part of the proof.
			\item \textit{Counter-example for $\Lambda\geq t_a+\frac{(t_p+1)r}{t_p}$}: Having established that $I\not=\emptyset$ for $\Lambda < t_a +\frac{(t_p+1)r}{t_p}$, we now demonstrate that for any $\Lambda\geq t_a+\frac{(t_p+1)r}{t_p}$, there always exist mini-subfiles such that $I=\emptyset$.

			Consider $\Lambda=t_a+r+\frac{r}{t_p}$. Since $\Lambda,t_a$ and $r$ are integers, it follows that $t_p$ divides $r$.

			Partition the set $[\Lambda]\setminus{S}$ into $(t_p+1)$ subsets $P_{i,i\in[t_p+1]}
			$ of cardinality $\frac{\Lambda-t_a}{t_p+1}=\frac{r}{t_p}$. For each $P_i$, define $Q_i=[\Lambda]\setminus\{S\cup P_i\}$. Note that $|Q_i|=\Lambda-t_a-\frac{\Lambda-t_a}{t_p+1}=\frac{(\Lambda-t_a)t_p}{t_p+1}=r.$ Since the sets $Q_{i,i\in[t_p+1]}$ are sets of cardinality $r$, they serve as placeholders for the user-index set ${U}$ and the mini-subfile-index sets ${T}_{i,i\in[t_p]}$. 
			
			Consider an element 
			$e\in[\Lambda]\setminus{S}$. Since the sets $P_{i,i\in[t_p+1]}$ partition the set $[\Lambda]\setminus{S}$, we know that $e$ belongs in one $P_i$, that is, $e\in P_i$ for some $i$, say $i^{\textasteriskcentered}$. Thus, we have $e\in \{Q_i:\forall i\in[t_p+1]\setminus \{i^{\textasteriskcentered}\}\}$. Hence, there exists $Q_{i^{\textasteriskcentered}}$ 
		such that 
		$e\not\in Q_{i^{\textasteriskcentered}}$. 
		Therefore, we have $\bigcap\limits_{i=1}^{t_p+1}Q_i=\emptyset$. 
		
		Thus, there will always exist a mini-subfile $W_{d_{{U}},{S},{T}_{i,i\in[t_p]}}$ such that $I=\{\bigcap\limits_{i=1}^{t_p}{T}_i\}\cap U=\emptyset$.
	\end{enumerate}
	\end{IEEEproof}
	\begin{thm}
		\label{intersectionclassrate}
		For a $(\Lambda,r,M_a,M_p,N)-$CMAP coded caching system such that $r>\frac{\Lambda(1-\frac{M_a}{N})}{1+\frac{N}{K M_p}}$, a worst-case rate 
		\begin{align}
			R= \sum\limits_{i=1}^{r-1} \frac{\binom{\Lambda-t_a}{i} \sum\limits_{j=0}^{r-i-1} (-1)^j \binom{\Lambda-t_a-i}{j} \binom{\binom{\Lambda-t_a-i-j}{r-i-j}}{t_p+1}}{\binom{\binom{\Lambda-t_a}{r}}{t_p} \binom{t_a+i}{t_a}},
		\end{align}is achievable for the subpacketization $F=\binom{\Lambda}{t_a}\binom{\binom{\Lambda-t_a}{r}}{t_p}$, where $t_a=\frac{\Lambda M_a}{N}\in[0,\Lambda]$ and $t_p=\frac{K M_p}{N}\in[0,K]$.
	\end{thm}
	\begin{IEEEproof}
		Section~\ref{achievability} presents the coded caching scheme that achieves the rate stated in Theorem~\ref{intersectionclassrate}. 
	\end{IEEEproof}\begin{remark}
		Note that while the rate in Theorem \ref{intersectionclassrate} is defined for $t_a,t_p\in\mathbb{Z}^+$, the lower convex envelope of the points in Theorem \ref{intersectionclassrate} is achievable via memory sharing.
	\end{remark}Observe that $r>\frac{\Lambda(1-\frac{M_a}{N})}{1+\frac{N}{K M_p}}$ can also be written as $\Lambda<t_a+r+\frac{r}{t_p}$. Moving forward, we will use this form of the condition. 
	\begin{defn}
		\label{uniformintersectiondefn}
		A $(\Lambda,r,M_a,M_p,N)-$CMAP coded caching system belongs to the uniform-intersection subclass if the cardinality of the intersection set $I$ remains constant across all mini-subfiles. 
	\end{defn}
	\begin{claim}
		\label{uniformintersectionclaim}
		A $(\Lambda, r, M_a, M_p, N)-$CMAP coded caching system belongs to the uniform-intersection subclass if the following two conditions hold: (i) $\binom{r}{i} \binom{\binom{\Lambda - t_a - i}{r - i} - 1}{t_p} = \binom{\binom{\Lambda - t_a}{r} - 1}{t_p}$, and (ii) $\binom{\Lambda - t_a - i - 1}{r - i - 1} < t_p + 1 \leq \binom{\Lambda - t_a - i}{r - i}$, where $|I| = i$, $i \in [1, r - 1]$, $t_a = \frac{\Lambda M_a}{N} \in [0, \Lambda]$, and $t_p = \frac{K M_p}{N} \in [0, K]$.
	\end{claim}
	\begin{IEEEproof}
		The proof is provided in Section \ref{uniformintersectionclaimproof}.
	\end{IEEEproof}
	\begin{cor}
		\label{uniformintersectionrate}
		For a $(\Lambda,r,M_a,M_p,N)-$CMAP coded caching system satisfying the conditions in Claim \ref{uniformintersectionclaim}, the worst-case rate $R=\frac{\binom{\Lambda-t_a}{r}-t_p}{(t_p+1)\binom{t_a+i}{t_a}},$ is achievable, with a subpacketization level $F=\binom{\Lambda}{t_a}\binom{\binom{\Lambda-t_a}{r}}{t_p}$, where $t_a=\frac{\Lambda M_a}{N}\in[0,\Lambda]$ and $t_p=\frac{K M_p}{N}\in[0,K]$.
	\end{cor}\begin{IEEEproof}
		The proof is presented in Section \ref{uniformintersectionrateproof}.
	\end{IEEEproof}
	\begin{thm}(Alpha Bound)
		\label{alphaboundrate}For a $(\Lambda,r, M_a, M_p,N)-$CMAP coded caching system and the placement policy described in Section \ref{achievability}, the optimal worst-case rate under uncoded placement is lower bounded as \begin{align}&R^{\textasteriskcentered}_{UC}(M_a,M_p)\geq\nonumber\\&\frac{\sum\limits_{i=1}^{\Lambda-t_a-r+1} \binom{\Lambda-r-i+1}{t_a}\binom{\binom{\Lambda-t_a}{r}-i}{t_p}+\sum\limits_{m=\Lambda-t_a-r+2}^{\binom{\Lambda-t_a}{r}} \binom{\binom{\Lambda-t_a}{r}-m}{t_p}}{\binom{\Lambda}{t_a}\binom{\binom{\Lambda-t_a}{r}}{t_p}},\end{align} where $t_a=\frac{\Lambda M_a}{N}\in[0,\Lambda]$ and $t_p=\frac{KM_p}{N}\in[0,K].$
	\end{thm}
	\begin{IEEEproof}
		The proof is presented in Section \ref{alphaboundrateproof}.
	\end{IEEEproof}
	
	While the lower bound presented in Theorem \ref{alphaboundrate} holds for any general $(\Lambda,r,M_a,M_p,N)-$CMAP coded caching system, we apply this lower bound as a baseline for evaluating the performance of the CMAP system considered in this work.

	\section{Achievability Scheme and Lower Bound}
	\label{achievabilitysection}
In this section, we first describe the placement and delivery scheme that achieves the rate stated in Theorem~\ref{intersectionclassrate}. We then present the proofs of Claim~\ref{uniformintersectionclaim} and Corollary~\ref{uniformintersectionrate}. Finally, we provide the proof of Theorem~\ref{alphaboundrate}.
	\subsection{Achievability Scheme}
	\label{achievability}
	Before presenting the delivery scheme, we introduce the functions used to construct the transmissions in the delivery scheme. We illustrate their use and the underlying intuition behind the proposed scheme through an example.
	\begin{defn}
		For the mini-subfile $W_{d_{{U}},{S},{T}_{i,i\in[t_p]}}$, the function $flip(W_{d_{{U}},{S},{T}_{i,i\in[t_p]}})$ is defined as $flip(W_{d_{{U}},{S},{T}_{i,i\in[t_p]}})=W_{d_{{U}},{S},{T}_{i,i\in[t_p]}}\bigoplus\limits_{j\in[t_p]} W_{d_{{T}_{j}},{S},\{{U},{T}_{i,i\in[t_p]\setminus\{j\}}\}}$\end{defn}
	\begin{remark} The $flip$ of a coded combination of mini-subfiles is defined as the coded combination of the $flip$ of the individual mini-subfiles. This is written below formally as $flip\bigl(\bigoplus\limits_{i=1}^{n} W_{d_{{U}^i},{S}^i,{T}^i_{j,j\in[t_p]}}\bigr)=\bigoplus\limits_{i=1}^{n} flip(W_{d_{{U}^i},{S}^i,{T}^i_{j,j\in[t_p]}}).$
	\end{remark}
	\begin{defn}
		For the mini-subfile $W_{d_{{U}},{S},{T}_{i,i\in[t_p]}}$, such that the intersection set $I={{U}}\bigcap\limits_{j\in[t_p]}{T}_j\not=\emptyset$, we define the function $swap_o(W_{d_{{U}},{S},{T}_{j,j\in[t_p]}},i)$ as $swap_o(W_{d_{{U}},{S},{T}_{j,j\in[t_p]}},i)=\nonumber\bigoplus\limits_{\substack{\widetilde{{U}}\subset I,|\widetilde{{U}}|=i,\widetilde{{S}}\subset S,|\widetilde{{S}}|=i}} W_{d_{\{{U}\cup\widetilde{{S}}\}\setminus\widetilde{{U}}},\{{S}\cup\widetilde{{U}}\}\setminus\widetilde{{S}},\{\{\{{T}_j\cup\widetilde{{S}}\}\setminus\widetilde{{U}}\}_{j\in[t_p]}\}}$, where $i\in[\min\{t_a,|I|\}]$.
	\end{defn}
	We now present an example that illustrates the proposed scheme and demonstrates the role of the functions defined above. Following the convention in \cite{SMR}, we write the user-index set $U$, subfile-index set $S$, and mini-subfile-index sets $T_{i,i\in[t_p]}$ without explicit set notation.
	\begin{example}
		\label{exampleintersectionclass}
		Consider a CMAP coded caching system in which a central server stores a library of $N = 10$ files. The system has $\Lambda = 5$ access caches, each with a memory of $M_a = 2$ files. Users connect to distinct subsets of $r = 3$ access caches, resulting in $K = \binom{\Lambda}{r} = 10$ users. Each user is also equipped with a private cache of size $M_p = 2$ files. The users are indexed by the subsets of access caches they connect to, denoted as $\{123,124,125,134,135,145,234,235,245,345\}$. In this setting, we have $t_a = \frac{\Lambda M_a}{N} = 1$. Each file is divided into $\binom{\Lambda}{t_a} = 5$ subfiles, indexed as $W_n = \{W_{n,i}:i \in [\Lambda]\}$. The contents of access cache $i$ are given by $Z_i = \{W_{n,i},\forall n \in [N]\}$. For example, 
			$Z_1=\{W_{n,1},\forall n\in[10]\}.$ 
		Each access cache stores $\frac{10}{5}=2$ files, satisfying its memory constraint. For this system, we have $t_p=\frac{K M_p}{N}=2$ and the server breaks each subfile $W_{n,{S}}$ into $\binom{\binom{\Lambda-t_a}{r}}{t_p}=6$ mini-subfiles as $W_{n,{S}}=\{W_{n,S,{T}_{i\in[t_p]}}:{T}_i\subset[\Lambda]\setminus {S},|{T}_i|=r,\forall i\in[t_p]\}$. For example, the subfile $W_{n,1}$ is broken down as $W_{n,1}=\bigl\{W_{n,1,234,235},W_{n,1,234,245},W_{n,1,234,345},W_{n,1,235,245},\\W_{n,1,235,345},W_{n,1,245,345}\bigr\}$. The server populates the private cache of users with the mini-subfiles of the subfiles it does not receive on connecting to the access caches. The contents of the private cache $Z^p_U$ are given as $Z^p_U=\{W_{n,{S},{T}_{i,i\in[t_p]}}:{S}\subseteq[\Lambda]\setminus{U},{T}_1={U},\{{T}_i\in\{{T}^\prime\subseteq[\Lambda]\setminus{S},|{T}^\prime|=r\},\forall i\in[2,t_p]\},\forall n\in[N]\}.$ For example, 
			$Z^p_{123}=\{W_{n,4,123,125},W_{n,4,123,135},W_{n,4,123,235},W_{n,5,123,124},\\W_{n,5,123,134},W_{n,5,123,234},\forall n\in[10]\}$. 
		Each private cache stores $\frac{6\times10}{6\times5}=2$ files, satisfying its memory constraint. Note that $t_a+r+\frac{r}{t_p}=1+3+\frac{3}{2}=5.5$ which is greater than $\Lambda=5$.
		
		%
		%
		We now explain how the transmissions in the delivery scheme are constructed. For a demand vector $\mathbf{d}=(d_{123},d_{124},d_{125},d_{134},d_{135},d_{145},d_{234},d_{235},d_{245},d_{345})$, consider a mini-subfile $W_{d_{123},4,125,135}$ demanded by the user $123$. 
		The intersection set $I=\{123\cap125\cap135\}=\{1\}$ for this mini-subfile. The server employs the function $swap_{o}(W_{d_{123},4,125,135},1)$ to generate $W_{d_{234},1,245,345}$. It then utilizes the function $flip(W_{d_{123},4,125,135}\oplus W_{d_{234},1,245,345})$ to generate the transmission $W_{d_{123},4,125,135}\oplus W_{d_{125},4,123,135}\oplus W_{d_{135},4,123,125}\oplus W_{d_{234},1,245,345}\oplus W_{d_{245},1,234,345}\oplus W_{d_{345},1,234,245}.$ The mini-subfiles $W_{d_{234},1,245,345}, W_{d_{245},1,234,345},$ and $W_{d_{345},1,234,245}$ are present with the user $123$ via $Z_1$. Further, the mini-subfiles $W_{d_{125},4,123,135}$ and $W_{d_{135},4,123,125}$ are stored in its private cache. Hence, the user is able to decode the transmission. Similar argument holds for the other users in the above transmission. 
		All transmissions made by the server are:
		\begin{enumerate}
			\item $W_{d_{123},4,125,135}\oplus W_{d_{125},4,123,135}\oplus W_{d_{135},4,123,125}\oplus W_{d_{234},1,245,345}\oplus W_{d_{245},1,234,345}\oplus W_{d_{345},1,234,245}$
			\item $W_{d_{123},4,125,235}\oplus W_{d_{125},4,123,235}\oplus W_{d_{235},4,123,125}\oplus W_{d_{134},2,145,345}\oplus W_{d_{145},2,134,345}\oplus W_{d_{345},2,134,145}$
			\item $W_{d_{123},4,135,235}\oplus W_{d_{135},4,123,235}\oplus W_{d_{235},4,123,235}\oplus W_{d_{124},3,145,245}\oplus W_{d_{145},3,124,245}\oplus W_{d_{245},3,124,145}$
			\item $W_{d_{123},5,124,134}\oplus W_{d_{124},5,123,134}\oplus W_{d_{134},5,123,124}\oplus W_{d_{235},1,245,345}\oplus W_{d_{245},1,235,345}\oplus W_{d_{345},1,235,245}$
			\item $W_{d_{123},5,124,234}\oplus W_{d_{124},5,123,234}\oplus W_{d_{234},5,123,124}\oplus W_{d_{135},2,145,345}\oplus W_{d_{145},2,135,345}\oplus W_{d_{345},2,135,145}$
			\item $W_{d_{123},5,134,234}\oplus W_{d_{134},5,123,234}\oplus W_{d_{234},5,123,134}\oplus W_{d_{125},3,145,245}\oplus W_{d_{145},3,125,245}\oplus W_{d_{245},3,125,145}$
			\item $W_{d_{124},3,125,145}\oplus W_{d_{125},3,124,145}\oplus W_{d_{145},3,124,125}\oplus W_{d_{234},1,235,345}\oplus W_{d_{235},1,234,345}\oplus W_{d_{345},1,234,235}$
			\item $W_{d_{124},3,125,245}\oplus W_{d_{125},3,124,245}\oplus W_{d_{245},3,124,125}\oplus W_{d_{134},2,135,345}\oplus W_{d_{135},2,134,345}\oplus W_{d_{345},2,134,135}$
			\item $W_{d_{124},5,134,234}\oplus W_{d_{134},5,124,234}\oplus W_{d_{234},5,124,134}\oplus W_{d_{125},4,135,235}\oplus W_{d_{135},4,125,235}\oplus W_{d_{235},4,125,135}$
			\item $W_{d_{134},2,135,145}\oplus W_{d_{135},2,134,145}\oplus W_{d_{145},2,134,135}\oplus W_{d_{234},1,235,245}\oplus W_{d_{235},1,234,245}\oplus W_{d_{245},1,234,235}$
		\end{enumerate}Each transmission is decodable by the intended users, and all demanded mini-subfiles are covered without repetition. As subpacketization $F = 5 \times 6 = 30$ and $10$ transmissions are made, the resulting rate is $R = \frac{10}{30} = 0.333$.
	\end{example}
	\begin{remark}
		In the above example, Claim~\ref{uniformintersectionclaim} is satisfied, ensuring that the intersection set of each mini-subfile has cardinality one. This results in a regular coded caching scheme where each transmission benefits an equal number of users. Users demand a total of 60 mini-subfiles and using Theorem~\ref{alphaboundrate}, we know the server needs to transmit at least $\frac{7}{30}$ files, that is, $7$ transmissions. In contrast, a regular coded caching scheme, where each transmission benefits an equal number of users, would require at least $10$ transmissions, as $10$ is the smallest integer greater than $7$ that divides $60$, demonstrating that the scheme in Example~\ref{exampleintersectionclass} is optimal under uncoded placement and regular transmission.
	\end{remark}
	We now give the general description of the placement and delivery phase of the proposed scheme.
	
	\textit{Placement Phase:} 
	Each file is first divided into $\binom{\Lambda}{t_a}$ non-overlapping subfiles of equal size as $W_n = \{W_{n,S} : S \subseteq [\Lambda], |S| = t_a\}$, where $t_a = \frac{\Lambda M_a}{N}$ represents the access cache replication factor. The contents of access cache $i$ are given by $Z_i = \{W_{n,S} : i \in S, S \subseteq [\Lambda], |S| = t_a, \forall n \in [N]\}$ for $i \in [\Lambda]$. Each access cache stores $\frac{N \binom{\Lambda - 1}{t_a - 1}}{\binom{\Lambda}{t_a}} = M_a$ files, satisfying its memory constraint.
	
	We will now describe how the private cache of the users are populated. Each subfile is further split into $\binom{\binom{\Lambda-t_a}{r}}{t_p}$ mini-subfiles as $W_{n,S}=\{W_{n,S,T_{i,i\in[t_p]}}:T_{i}\subset[\Lambda]\setminus S,|T_i|=r,\forall i\in[t_p]\}$, where $t_p=\frac{KM_p}{N}$ is the private cache memory replication factor. We denote the mini-subfile of subfile $S$ of the file $n$ as $W_{n,{S},{T}_{i,i\in[t_p]}}$ where ${T}_{i,i\in[t_p]}$ are the users possessing the mini-subfile $W_{n,{S},{T}_{i,i\in[t_p]}}$ in their private cache. The contents of the private cache of user ${U}$ are $Z^p_{{U}}=\{W_{n,{S},{T}_{i,i\in[t_p]}}:{S}\subseteq[\Lambda]\setminus{U},{T}_1={U},\{{T}_i\in\{{T}^\prime\subseteq[\Lambda]\setminus{S},|{T}^\prime|=r\},\forall i\in[2,t_p]\},\forall n\in[N]\}.$ The private cache of each user stores $\frac{N\binom{\Lambda-r}{t_a}\binom{\binom{\Lambda-t_a}{r}-1}{t_p-1}}{\binom{\Lambda}{t_a}\binom{\binom{\Lambda-ta}{r}}{t_p}}=M_p$ files, satisfying its memory constraint. As per the proposed placement policy, there is no overlap between the contents of a user's private cache and the contents of the access caches the user connects to. Notably, for $t_p=1$, this placement reduces to the one discussed in \cite{SMR}.
	
	\textit{Delivery Phase:} We define the user-demand set ${D}_{U}$ of user $U$ as $D_U=\{(S,T_{i,i\in[t_p]}):S\subseteq[\Lambda],|S|=t_a,T_i\subseteq[\Lambda],|T_i|=r,T_i\not=U,\forall i\in[t_p]\}$. In other words, the user-demand set $D_U$ contains the indices of all the mini-subfiles demanded by user $U$. For an element $({S},{T}_{i,i\in[t_p]})\in D_U$, the server transmits $flip\bigl(W_{d_{{U}},{S},{T}_{l,l\in[t_p]}}\bigoplus\limits_{k\in[\min\{|I|,t_a\}]} swap_o\bigl(W_{d_{{U}},{S},{T}_{l,l\in[t_p]}},k\bigr)\bigr),$ where $I=\{\bigcap\limits_{i=1}^{t_p}{T}_i\}\cap U$. After each transmission, the server removes the indices of the mini-subfiles involved in the transmission from the user-demand sets of the respective users. The server continues making transmissions until the user-demand set of all users are empty, that is, $D_U=\emptyset,\forall U\subset[\Lambda],|U|=r$.
	
	
	\textit{Decodability}: For an element $({S},{T}_{i,i\in[t_p]})$ in the user-demand set ${D}_{{U}}$ of the user ${U}$, consider the transmission $flip\bigl(W_{d_{{U}},{S},{T}_{l,l\in[t_p]}}\bigoplus\limits_{k\in[\min\{|I|,t_a\}]} swap_o\bigl(W_{d_{{U}},{S},{T}_{l,l\in[t_p]}},k\bigr)\bigr)$. We will now explain how the user ${U}$ decodes this transmission. Observe that the above transmission can be written as $flip\bigl(W_{d_{{U}},{S},{T}_{l,l\in[t_p]}}\bigr)\bigoplus\limits_{k\in[\min\{|I|,t_a\}]}\\flip\bigl( swap_o\bigl(W_{d_{{U}},{S},{T}_{l,l\in[t_p]}},k\bigr)\bigr)$. 
	Consider all the mini-subfiles present in $\bigoplus\limits_{k\in[\min\{|I|,t_a\}]}flip\bigl( swap_o\bigl(W_{d_{{U}},{S},{T}_{l,l\in[t_p]}},k\bigr)\bigr)$. Since $k$ elements of the set $I=\{\bigcap\limits_{i=1}^{t_p} {T}_i\}\cap U$ are swapped with $k$ elements of the set ${S}$, where $k\in[\min\{|I|,t_a\}]$, these mini-subfiles are available to user ${U}$. This is due to the placement policy of the access caches, where a user ${U}$ has access to the subfile (and consequently all the mini-subfiles of the subfile) $W_{n,{S}}$ if ${S}\cap{U}\not=\emptyset$.
	
	Next, consider the mini-subfiles present in $flip\bigl(W_{d_{{U}},{S},{T}_{l,l\in[t_p]}}\bigr)$. The user ${U}$ has all the mini-subfiles present in $flip\bigl(W_{d_{{U}},{S},{T}_{l,l\in[t_p]}}\bigr)$
	, except the mini-subfile $W_{d_{{U}},{S},{T}_{l,l\in[t_p]}}$. This is because all the other mini-subfiles contain ${U}$ in their mini-subfile-index sets. Thus, it follows from the placement policy of the private caches that user ${U}$ has access to these mini-subfiles. 
	
	
	Thus, user $U$ can cancel all interfering mini-subfiles to decode its desired mini-subfile. The same holds for all users in the transmission. Since the server continues transmitting until all demand sets are empty and every user can decode the transmissions to obtain their desired mini-subfiles, all the demands of the users are satisfied.
	
	\textit{Rate Characterization:} 
	\label{ratecharacterization}
	We first characterize the total number of demanded mini-subfiles $W_{d_{U},S,T_{i,i\in[t_p]}}$ having an intersection set $ I=\{\bigcap\limits_{j=1}^{t_p} T_j\}\cap U$ of cardinality $ i $, where $i\in[1,r-1]$. 
	
	Consider a specific subfile-index set $ {S} $ of cardinality $ t_a $ and an intersection set $ I $ of cardinality $ i $. The problem of determining the number of mini-subfiles of a given subfile $ {S} $ having an intersection set $I$ of cardinality $ |I| = i $ reduces to identifying $ (t_p+1) $ sets $ A_{j,j\in[t_p+1]}$ of cardinality $ r $ such that their intersection satisfies $ \bigcap\limits_{j=1}^{t_p+1} A_j = I $. Once these sets are identified, one set is chosen as the user-index set $U$ and the remaining sets will form the mini-subfile-index sets $T_{j,j\in[t_p]}$. Moving forward, we denote the cardinality of the intersection $I$ as $|I|$. We now outline the procedure for calculating the number of collections of such sets. 

	To obtain the number of collections of $(t_p+1)$ sets $A_{j,j \in [t_p]}$ of cardinality $r$ whose intersection is exactly equal to a given set $I$, we begin by counting the number of such collections where each set contains all elements of $I$. Out of the set $[\Lambda]$, we remove $t_a$ elements for the subfile-index set $S$ and $|I|$ elements are reserved in every set $A_j$ to ensure an intersection set of $I$. From the remaining elements, we pick $r - |I|$ elements to form candidate sets, from which we choose $(t_p + 1)$ sets, yielding \( \binom{\binom{\Lambda - t_a - |I|}{r - |I|}}{t_p + 1} \) such collections. However, the intersection of these sets may be a superset of $I$, since the sets may share additional elements beyond those in $I$. To correct for this and count only those collections whose intersection is exactly $I$, we apply the Principle of Inclusion-Exclusion (PIE). This method alternately subtracts and adds the number of collections where the sets contain additional common elements, leading to $\sum\limits_{j=0}^{r-|I|-1} (-1)^j \binom{\Lambda-t_a-|I|}{j} \binom{\binom{\Lambda-t_a-|I|-j}{r-|I|-j}}{t_p+1}$ collections of $(t_p+1)$ sets having an intersection set $I$.
	
	To determine the total number of mini-subfiles corresponding to these $ (t_p+1) $ sets, we designate one among them as the user-index set $U$, and the remaining $ t_p $ sets form the mini-subfile-index sets $T_{i,i\in[t_p]}$. Since the selection of a user-index set $U$ can be made in $ (t_p+1) $ ways, the number of mini-subfiles of a particular subfile associated with the intersection set $ I $ is $(t_p+1) \sum\limits_{j=0}^{r-|I|-1} (-1)^j \binom{\Lambda-t_a-|I|}{j} \binom{\binom{\Lambda-t_a-|I|-j}{r-|I|-j}}{t_p+1}.$
	
	Finally, the total number of mini-subfiles corresponding to any subfile of cardinality $ t_a $ and an intersection set $I$ of cardinality $ |I| $ is $\binom{\Lambda}{t_a} (t_p+1) \binom{\Lambda-t_a}{|I|} \sum\limits_{j=0}^{r-|I|-1} (-1)^j \binom{\Lambda-t_a-|I|}{j} \binom{\binom{\Lambda-t_a-|I|-j}{r-|I|-j}}{t_p+1}.$ Here, the terms $ \binom{\Lambda}{t_a} $ and $ \binom{\Lambda-t_a}{|I|} $ account for the $ \binom{\Lambda}{t_a} $ possible subfiles and $ \binom{\Lambda-t_a}{|I|} $ possible subsets of cardinality $ |I| $ of the set $ [\Lambda] \setminus {S} $. 
	
	Having determined the number of mini-subfiles that yield an intersection set of cardinality $|I|$, we now calculate the number of such mini-subfiles included in a transmission. 
	
	For an intersection set of cardinality $ |I| $, each transmission is a coded combination of $(t_p+1) \sum\limits_{l=0}^{\min\{|I|,t_a\}} \binom{|I|}{l} \binom{t_a}{l}$ mini-subfiles. This is because the function $swap_o$ exchanges $l$ elements of the intersection set $I=\{\bigcap\limits_{k=1}^{t_p} T_k\}\cap U$ with $l$ elements of the subfile-index set $S$, where $|S|=t_a$ and $l\in[min\{|I|,t_a\}]$. Further, the function $flip$ exchanges the user-index set with one of the mini-subfile-index sets, generating $(t_p+1)$ mini-subfiles for each mini-subfile generated by the function $swap_o$. Using the Vandermonde's identity, we simplify $(t_p+1) \sum\limits_{l=0}^{\min\{|I|,t_a\}} \binom{|I|}{l} \binom{t_a}{l}$ to $(t_p+1) \binom{t_a+|I|}{t_a}$. As the cardinality of the intersection set $I$ can range from $1$ to $r-1$, we have the number of transmissions as $\sum\limits_{|I|=1}^{r-1}\frac{\text{Number of mini-subfiles having } |I|}{(t_p+1) \binom{t_a+|I|}{t_a}}$ and the rate then becomes $\frac{\text{Number of transmissions}}{\binom{\Lambda}{t_a}\binom{\binom{\Lambda-t_a}{r}}{t_p}}$. Therefore, we have $R=\sum\limits_{|I|=1}^{r-1} \frac{\binom{\Lambda}{t_a} (t_p+1) \binom{\Lambda-t_a}{|I|} \sum\limits_{j=0}^{r-|I|-1} (-1)^j \binom{\Lambda-t_a-|I|}{j} \binom{\binom{\Lambda-t_a-|I|-j}{r-|I|-j}}{t_p+1}}{\binom{\Lambda}{t_a} \binom{\binom{\Lambda-t_a}{r}}{t_p} (t_p+1) \binom{t_a+|I|}{t_a}}$, which simplifies to 
	\begin{align}
		R= \sum\limits_{|I|=1}^{r-1} \frac{\binom{\Lambda-t_a}{|I|} \sum\limits_{j=0}^{r-|I|-1} (-1)^j \binom{\Lambda-t_a-|I|}{j} \binom{\binom{\Lambda-t_a-|I|-j}{r-|I|-j}}{t_p+1}}{\binom{\binom{\Lambda-t_a}{r}}{t_p} \binom{t_a+|I|}{t_a}}.
	\end{align}
	
	\subsection{Uniform-Intersection Subclass}
	In this subsection, we first provide the proof of Claim~\ref{uniformintersectionclaim}, which establishes conditions on the system parameters under which a CMAP coded caching system belongs to the uniform-intersection subclass. We then characterize the rate performance of this subclass, as stated in Corollary~\ref{uniformintersectionrate}.
	
	Consider the following proof of Claim \ref{uniformintersectionclaim}.
	
	\label{uniformintersectionclaimproof}
	\begin{IEEEproof}
		We begin the proof by defining ${N}_k$ and $C_k$. Let ${N}_k=\binom{\Lambda}{t_a}\binom{\Lambda-t_a}{k}(t_p+1)\binom{\binom{\Lambda-t_a-k}{r-k}}{t_p+1}$ denote the number of demanded mini-subfiles having an intersection set of cardinality atleast $k$. Further, let $C_k=\binom{\Lambda}{t_a} (t_p+1) \binom{\Lambda-t_a}{k} \sum\limits_{j=0}^{r-k-1} (-1)^j \binom{\Lambda-t_a-k}{j} \binom{\binom{\Lambda-t_a-k-j}{r-k-j}}{t_p+1}$ denote the number of demanded mini-subfiles having an intersection set of cardinality exactly $k$. Observe that $C_k$ can be written as:
		\begin{align*}
			C_k=\sum\limits_{j=0}^{r-k-1} (-1)^j \binom{\Lambda-t_a-k}{j} {N}_{k+j}.
		\end{align*}Since in the uniform-intersection subclass, all mini-subfiles have an intersection set of cardinality $|I|$, it follows that ${N}_{k}=0,k\in[|I|+1,r-1]$. Therefore, we have 
		\begin{align}
			C_{|I|}&={N}_{|I|}+\sum\limits_{j=1}^{r-k-1} (-1)^j \binom{\Lambda-t_a-k}{j} {N}_{|I|+j}\nonumber\nonumber\\\label{ci1}&=\binom{\Lambda}{t_a} (t_p+1) \binom{\Lambda-t_a}{|I|}\binom{\binom{\Lambda-t_a-|I|}{r-|I|}}{t_p+1}.
		\end{align}Since ${N}_{|I|}\not=0,$ it follows that $t_p+1\leq {\binom{\Lambda-t_a-|I|}{r-|I|}}$. However as ${N}_{k}=0,k\in[|I|+1,r-1]$, we have $t_p+1>{\binom{\Lambda-t_a-|I|-1}{r-|I|-1}}$, yielding the second condition. Further, as every mini-subfile has an intersection set of cardinality $|I|$, the total number of demanded mini-subfiles equals $C_{|I|}$. Thus, we have \begin{align}
			\label{ci2}C_{|I|}=\binom{\Lambda}{r}\binom{\Lambda-r}{t_a}\binom{\binom{\Lambda-t_a}{r}-1}{t_p}.
		\end{align}Equating equations \eqref{ci1} and \eqref{ci2}, we have $\binom{\Lambda}{t_a}\binom{\Lambda-t_a}{|I|}(t_p+1)\binom{\binom{\Lambda-t_a-|I|}{r-|I|}}{t_p+1}=\binom{\Lambda}{r}\binom{\Lambda-r}{t_a}\binom{\binom{\Lambda-t_a}{r}-1}{t_p}$. Using the property that $\binom{n}{k}\binom{n-k}{l}=\binom{n}{l}\binom{n-l}{k}$, we get $\binom{\Lambda-t_a}{|I|}(t_p+1)\binom{\binom{\Lambda-t_a-|I|}{r-|I|}}{t_p+1}=\binom{\Lambda-t_a}{r}\binom{\binom{\Lambda-t_a}{r}-1}{t_p}$. We obtain $\binom{\Lambda-t_a}{|I|}{\binom{\Lambda-t_a-|I|}{r-|I|}}\binom{\binom{\Lambda-t_a-|I|}{r-|I|}-1}{t_p}=\binom{\Lambda-t_a}{r}\binom{\binom{\Lambda-t_a}{r}-1}{t_p}$ using the property that $\binom{n+1}{k+1}=\frac{n+1}{k+1}\binom{n}{k}$. Finally, we simplify the condition to \begin{align*}\binom{r}{|I|}\binom{\binom{\Lambda-t_a-|I|}{r-|I|}-1}{t_p}=\binom{\binom{\Lambda-t_a}{r}-1}{t_p},\end{align*}using the property that $\binom{\Lambda-t_a}{i}{\binom{\Lambda-t_a-i}{r-i}}=\binom{r}{i}\binom{\Lambda-t_a}{r}.$
	\end{IEEEproof}
	
	 We now characterize the rate performance of the uniform-intersection subclass as given below.
	
	\begin{IEEEproof}
		Given that the conditions in Claim \ref{uniformintersectionclaim} are satisfied, all the demanded mini-subfiles have an intersection set of cardinality $|I|=i$. Thus, the rate becomes $R=\frac{\text{Number of mini-subfiles having $|I|=i$}}{\binom{\Lambda}{t_a}\binom{\binom{\Lambda-t_a}{r}}{t_p}(t_p+1)\binom{t_a+i}{t_a}}=\frac{\text{Number of demanded mini-subfiles}}{\binom{\Lambda}{t_a}\binom{\binom{\Lambda-t_a}{r}}{t_p}(t_p+1)\binom{t_a+i}{t_a}}$. Using this we have, 
		\label{uniformintersectionrateproof}
		\begin{align}
			R=&\frac{\binom{\Lambda}{r}\binom{\Lambda-r}{t_a}\binom{\binom{\Lambda-t_a}{r}-1}{t_p}}{\binom{\Lambda}{t_a}\binom{\binom{\Lambda-t_a}{r}}{t_p}(t_p+1)\binom{t_a+i}{t_a}}\nonumber\\
			R=&\frac{\binom{\Lambda-t_a}{r}\binom{\binom{\Lambda-t_a}{r}-1}{t_p}}{\binom{\binom{\Lambda-t_a}{r}}{t_p}(t_p+1)\binom{t_a+i}{t_a}}\nonumber\nonumber\\
			R=&\frac{\binom{\Lambda-t_a}{r}}{(t_p+1)\binom{t_a+i}{t_a}}\left(1-\frac{t_p}{\binom{\Lambda-t_a}{r}}\right)\nonumber\\
			R=&\frac{\binom{\Lambda-t_a}{r}-t_p}{(t_p+1)\binom{t_a+i}{t_a}},
		\end{align}
	\end{IEEEproof}

	\begin{remark}
		Observe that in the uniform-intersection subclass, the cardinality of the intersection set is constant across all mini-subfiles. Therefore, the resulting scheme is a regular coded caching scheme with a rate as specified in Corollary \ref{uniformintersectionrate}, and a coding gain $g=(t_p+1)\binom{t_a+i}{t_a}$, where the coding gain is defined as the total number of users benefiting from each transmission.
		
	\end{remark}
	
	It is important to note that each transmission in the proposed scheme benefits $(t_p+1)\binom{t_a+i}{t_a}$ users, where $i \in [1, r-1]$ represents the cardinality of the intersection set $I$. As the number of sets involved in the intersection increases, i.e., as $ t_p$ increases, the cardinality $i$ of the intersection set $I$ will decrease. Consequently, while the factor $(t_p+1)$ increases linearly with $t_p$, the term $\binom{t_a+i}{t_a}$ decreases exponentially as $i$ decreases with increasing $t_p$, leading to poorer performance for larger values of $t_p$.
	\subsection{Alpha Bound}
		\label{alphaboundrateproof}
%

In this subsection, we formulate the delivery phase of the scheme described in Subsection~\ref{achievability} as an ICP, following the approach in \cite{KTR}. Specifically, we identify a set of mini-subfiles requested by users such that no user demanding a mini-subfile in the set possesses any other mini-subfile in the set as side information; this set forms a generalized independence set. The cardinality of the largest such set, denoted by $\alpha(\mathcal{I})$, provides a lower bound on the number of transmissions required to satisfy all user demands. Dividing this bound by the subpacketization $F$ yields a lower bound on the optimal worst-case rate, as stated in Theorem~\ref{alphaboundrate}.
			
			We define ${U}_i$ as the $i^{th}$ user, $i\in\left[\binom{\Lambda}{r}\right]$, and ${U}^{{S}}_{l}$ as the $l^{th}$ user, demanding the subfile ${S}$, $l\in\left[\binom{\Lambda-t_a}{r}\right]$, under lexicographical ordering of the user-index sets. 
			
			We construct the set $B(\mathbf{d})=B_1(\mathbf{d})\cup B_2(\mathbf{d})$ whose elements are messages of the ICP $\mathcal{I}$ such that the set of indices of the messages in $B(\mathbf{d})$ forms a generalized independent set, where $B_1(\mathbf{d})=\bigcup\limits_{i\in[1,\Lambda-r-t_a+1]}\bigcup\limits_{L\subset\left[i+1,\binom{\Lambda-ta}{r}\right],|L|=t_p}\{W_{d_{{U}_i},{S},{U}^{{S}}_{{l,l\in L}}}, {S}\subset[r+i,\Lambda],|{S}|=t_a\},$ and $B_2(\mathbf{d})=\bigcup\limits_{m\in\left[\Lambda-r-ta+2,\binom{\Lambda-ta}{r}\right]}\bigcup\limits_{L\subset\left[m+1,\binom{\Lambda-t_a}{r}\right],|L|=t_p}\{W_{d_{{U}^{{S^{\prime}}}_m},{S}^{\prime},{U}^{{S}^{\prime}}_{l,l\in L }}\},$ where ${S}^{\prime}=[\Lambda+1-t_a,\Lambda]$. \begin{claim}
				The indices of the set $B(\mathbf{d})$ form a generalized independence set. 
			\end{claim}
			\begin{IEEEproof}
				Let the set of indices of $B(\mathbf{d})$ be $H(\mathbf{d})$. Observe that each message in $B(\mathbf{d})$ is wanted by one receiver. Hence, all subsets of $H(\mathbf{d})$ of cardinality one are present in $\mathcal{J}(\mathcal{I})$. Next, consider a set $C=\{W_{d_{{U}_{i_1}},{S}_{j_1},{U}^{{S}_{j_1}}_{l,l\in L_{k_1}}}, \cdots, W_{d_{{U}_{i_c}},{S}_{j_c},{U}^{{S}_{j_c}}_{l,l\in L_{k_c}}}\}\subseteq B(\mathbf{d}),$ where $i_1\leq i_2\leq\cdots\leq i_c$.

				Consider the message $W_{d_{{U}_{i_1}},{S}_{j_1},{U}^{{S}_{j_1}}_{l,l\in L_{k_1}}}$. Since the receiver demanding this
				message has no other message in $C$ as side information, the indices of messages in $C$ lie in $\mathcal{J}(\mathcal{I})$ and
				any subset of $H(\mathbf{d})$ will lie in $\mathcal{J}(\mathcal{I})$. Thus, $H(\mathbf{d})$ forms a generalized independence set.
			\end{IEEEproof}Since $H(\mathbf{d})$ forms a generalized independence set, it follows that $\alpha(\mathcal{I})\geq |H(\mathbf{d})|\implies\alpha(\mathcal{I})\geq |B(\mathbf{d})|$ as $|H(\mathbf{d})|=|B(\mathbf{d})|$. Since the elements of $B_1(\mathbf{d})$ and $B_2(\mathbf{d})$ are disjoint, we have $|B(\mathbf{d})|=|B_1(\mathbf{d})|+|B_2(\mathbf{d})|$. We will now calculate $|B_1(\mathbf{d})|$. 
			
			Observe that a user ${U}_i$ in the set $B_1(\mathbf{d})$ demands $\binom{\Lambda-r-i+1}{t_a}$ subfiles. Further, for each such subfile, user ${U}_i$ demands $\binom{\binom{\Lambda-t_a}{r}-i}{t_p}$ mini-subfiles. Therefore, it follows that $|B_1(\mathbf{d})|=\sum\limits_{i=1}^{\Lambda-t_a-r+1} \binom{\Lambda-r-i+1}{t_a}\binom{\binom{\Lambda-t_a}{r}-i}{t_p}$. We now compute $|B_2(\mathbf{d})|$.
			
			Each user ${U}^{{S}^{\prime}}_{m}$ in the set $B_2(\mathbf{d})$ demands $\binom{\binom{\Lambda-t_a}{r}-m}{t_p}$ mini-subfiles of the subfile ${S}^{\prime}$. Therefore, $|B_2(\mathbf{d})|=\sum\limits_{m=\Lambda-r-t_a+2}^{\binom{\Lambda-t_a}{r}} \binom{\binom{\Lambda-t_a}{r}-m}{t_p}.$
			
			Hence, we conclude that $|B(\mathbf{d})|=\sum\limits_{i=1}^{\Lambda-t_a-r+1} \binom{\Lambda-r-i+1}{t_a}\binom{\binom{\Lambda-t_a}{r}-i}{t_p}+\sum\limits_{m=\Lambda-r-t_a+2}^{\binom{\Lambda-t_a}{r}} \binom{\binom{\Lambda-t_a}{r}-m}{t_p}$. Since, $\alpha(\mathcal{I})\geq |B(\mathbf{d})$, the lower bound on the number of transmissions required to be made by the server in order to satisfy the demands of the users is as follows $T\geq \sum\limits_{i=1}^{\Lambda-t_a-r+1} \binom{\Lambda-r-i+1}{t_a}\binom{\binom{\Lambda-t_a}{r}-i}{t_p}+\sum\limits_{m=\Lambda-r-t_a+2}^{\binom{\Lambda-t_a}{r}} \binom{\binom{\Lambda-t_a}{r}-m}{t_p}.$
		For the given $\Lambda,r,t_a$ and $t_p$, the subpacketization is fixed at $F=\binom{\Lambda}{t_a}\binom{\binom{\Lambda-t_a}{r}}{t_p}$. Hence, a lower bound on the rate can be obtained by dividing the lower bound on the number of transmissions by the subpacketization. Therefore, dividing the lower bound on transmissions obtained in Theorem \ref{alphaboundrate}, we get $R^{\textasteriskcentered}_{UC}(M_a,M_p)=\frac{T}{\binom{\Lambda}{t_a}\binom{\binom{\Lambda-t_a}{r}}{t_p}}\geq\frac{\sum\limits_{i=1}^{\Lambda-t_a-r+1} \binom{\Lambda-r-i+1}{t_a}\binom{\binom{\Lambda-t_a}{r}-i}{t_p}+\sum\limits_{m=\Lambda-r-t_a+2}^{\binom{\Lambda-t_a}{r}} \binom{\binom{\Lambda-t_a}{r}-m}{t_p}}{\binom{\Lambda}{t_a}\binom{\binom{\Lambda-t_a}{r}}{t_p}}.$
	\section{Numerical Comparisons}
	\label{numericalcomparisons}
	\begin{figure}
		\centering
		\includegraphics[width=0.52\textwidth,height=0.25075\textwidth]{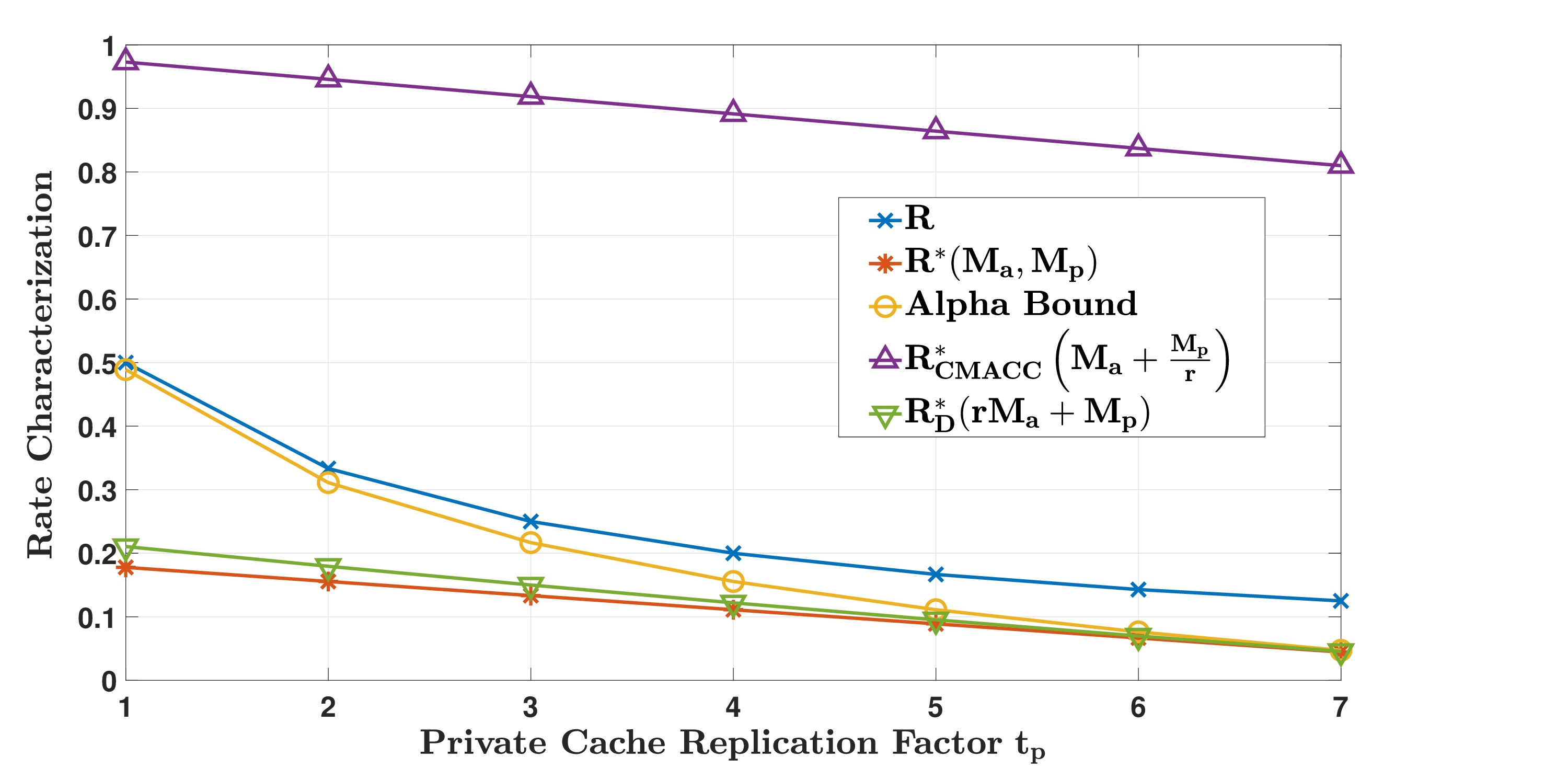}
		\caption{Rate vs $t_p$ for a $(\Lambda=10,r=8,M_a=4.5,M_p,N=45)-$CMAP Coded Caching system.}
		\label{subclassfig}
	\end{figure}
	\begin{figure}
		\centering
		\includegraphics[width=0.52\textwidth,height=0.25075\textwidth]{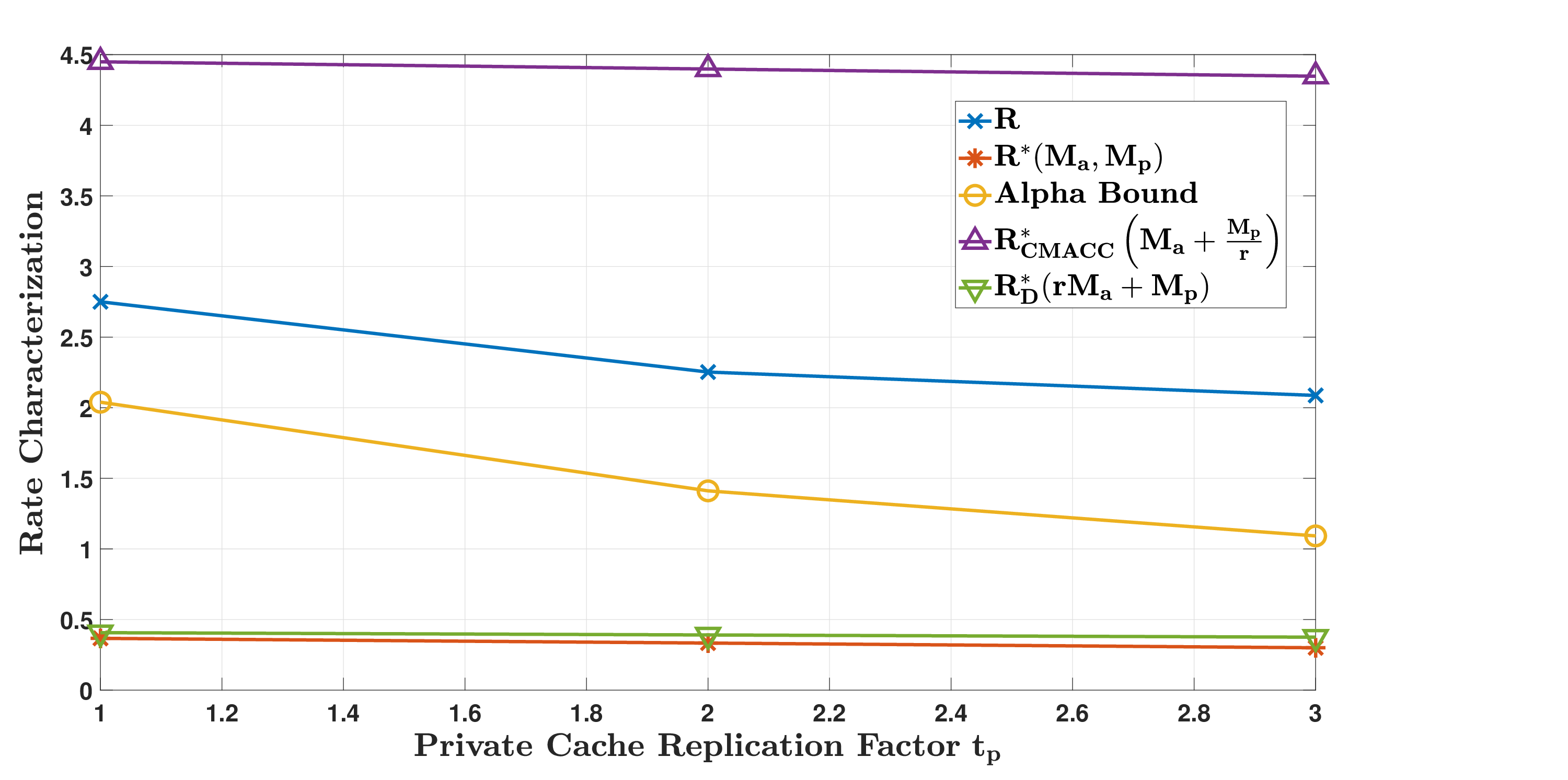}
		\caption{Rate vs $t_p$ for a $(\Lambda=10,r=7,M_a=12,M_p,N=120)-$CMAP Coded Caching system.}
		\label{classfig}
	\end{figure}

	This section presents numerical comparisons between the proposed scheme and the bounds derived in Theorem~\ref{alphaboundrate} and discussed in Section~\ref{Preliminarybounds}.
	
	Fig.~\ref{subclassfig} corresponds to a CMAP system with $\Lambda = 10$ access caches, each storing $M_a = 4.5$ files, and a library of $N = 45$ files. Each user accesses $r = 8$ caches, resulting in $K = 45$ users. The private cache size $M_p$ varies from $1$ to $7$. This system satisfies the condition in Claim \ref{uniformintersectionclaim}, placing it in the uniform-intersection subclass.
	
	Fig.~\ref{classfig} corresponds to a $(\Lambda = 10, r = 7, M_a = 12, M_p, N = 120)-$CMAP system. It has $\Lambda = 10$ access caches, each user accesses $r = 7$ of them, and each access cache stores $M_a = 12$ files. The library contains $N = 120$ files, and the private cache size $M_p$ ranges from $1$ to $3$. The number of users is $K = \binom{10}{7} = 120$. This configuration satisfies the condition for the intersection class but not for the uniform-intersection subclass.
	
	For both systems, we compare the rate of the proposed scheme with the CMACC scheme from \cite{SMR} (with cache memory $M_a + \frac{M_p}{r}$), the MAN scheme from \cite{MAN} (with cache memory $rM_a + M_p$), the index coding-based lower bound from Theorem~\ref{alphaboundrate}, and the cut-set bound from \cite{SMR}. It can be seen that the proposed scheme achieves rates close to the lower bound presented in Theorem \ref{alphaboundrate}.
	
	%
	%
	\section{Conclusions}
	In this work, we have explored a constrained indexing regime, referred to as the intersection class, and introduced a novel coded caching scheme while characterizing its rate-memory trade-off. We also examined the uniform-intersection subclass, a special case within the intersection class, where the cardinality of the intersection set remains constant across all mini-subfiles. This subclass leads to a regular coded caching scheme, for which we provide a rate characterization. Additionally, we derived an index coding-based lower bound on the achievable rate under uncoded placement. Numerical comparisons between our proposed scheme, the new lower bound, and bounds from the original work were presented.
	
		\section*{Acknowledgment}
		This work was supported partly by the Science and Engineering Research Board (SERB) of the Department of Science and Technology (DST), Government of India, through J.C Bose National Fellowship to Prof. B. Sundar Rajan.

\clearpage
\end{document}